\documentclass{article}

\usepackage{arxiv}

\usepackage{amsmath,amsfonts,amssymb}  
\usepackage{booktabs}               
\usepackage[inline]{enumitem}
\usepackage{algpseudocode}
\usepackage{algorithm}
\usepackage{array}
\usepackage[caption=false,font=small,labelfont=sf,textfont=sf]{subfig}
\usepackage{multirow}
\usepackage{textcomp}
\usepackage{stfloats}
\usepackage{url}
\usepackage{verbatim}
\usepackage{graphicx}
\graphicspath{{figures}}
\usepackage{cite}
\usepackage{appendix}
\newcommand{\secref}[1]{Section~\ref{#1}}

\newcommand{\figref}[1]{Fig.~\ref{#1}}                  
\newcommand{\figsref}[2]{Figs.~\ref{#1}--\ref{#2}}      
\newcommand{\Figref}[1]{Figure~\ref{#1}}                
\newcommand{\tabref}[1]{Table~\ref{#1}}                 
\newcommand{\Tabref}[1]{Table~\ref{#1}}                 
\newcommand{\equref}[1]{Eq.~(\ref{#1})}

\newcommand{\algoref}[1]{Algorithm~\ref{#1}}
\newcommand{\appref}[1]{Appendix~\ref{#1}}                 

\newcommand{\cf}{cf.\ }
\newcommand{\ie}{i.e.,\ }

\DeclareMathOperator*{\argmin}{\arg \min}
\DeclareMathOperator*{\argmax}{\arg \max}
\DeclareMathOperator*{\E}{\mathbb{E}}

\begin{document}

\title{PhysQ: A Physics Informed Reinforcement Learning Framework for Building Control}

\author{Gargya~Gokhale, %
Bert~Claessens, %
and %
Chris~Develder%
}

\maketitle

\begin{abstract}
Large-scale integration of intermittent renewable energy sources calls for substantial demand side flexibility. Given that the built environment accounts for approximately 40\% of total energy consumption in EU, unlocking its flexibility is a key step in the energy transition process. This paper focuses specifically on energy flexibility in residential buildings, leveraging their intrinsic thermal mass. Building on recent developments in the field of data-driven control, we propose PhysQ. As a physics-informed reinforcement learning framework for building control, PhysQ forms a step in bridging the gap between conventional model-based control and data-intensive control based on reinforcement learning. Through our experiments, we show that the proposed PhysQ framework can learn high quality control policies that outperform a business-as-usual, as well as a rudimentary model predictive controller. Our experiments indicate cost savings of about 9\% compared to a business-as-usual controller. Further, we show that PhysQ efficiently leverages prior physics knowledge to learn such policies using fewer training samples than conventional reinforcement learning approaches, making PhysQ a scalable alternative for use in residential buildings. Additionally, the PhysQ control policy utilizes building state representations that are intuitive and based on conventional building models, that leads to better interpretation of the learnt policy over other data-driven controllers.
\end{abstract}


\section{Introduction}
\label{sec:intro}
Global energy markets are facing unprecedented stress because of factors such as the post-pandemic re-opening of societies, geopolitical conflicts, and extreme climate events~\cite{mckinsey-report}. This has led to a steep increase in energy prices, with average households expecting a three-to-four-fold increase in their annual energy bills.
With these prices expected to persist, a possible approach to tackle this problem is to unlock demand side flexibility of residential buildings and exploit the high volatility of these prices~\cite{price-volatility,dr-sota-2022}. 

This paper focuses specifically on unlocking the potential flexibility of residential buildings for optimizing energy usage and consequently reducing energy bills.
Energy flexibility in residential buildings is typically characterized by the ability to shift heating/cooling energy for a period without jeopardizing the comfort of the user~\cite{building-flex}. 
A popular approach for exploiting this flexibility includes utilizing the building thermal mass --- an intrinsic, cost-free passive heat storage element of the building~\cite{thermal-mass}.
However, exploiting this flexibility necessitates solving a sequential decision making problem with partial observability and under uncertain operating conditions.

Previous works~\cite{mpc-review-2021,rl-review, bems-survey} have extensively reviewed different control algorithms for this task, most notably Model Predictive Control (MPC). MPC relies on a mathematical model of the system to anticipate its future behavior and an optimizer that uses this model to obtain control actions~\cite{mpc-basic}. Previous research directions include exploring different system modelling techniques~\cite{ctrl-oriented-models, building_model_crucial} and different optimization techniques for solving a receding horizon optimization problem~\cite{economic-mpc, stochastic-mpc, scenario-mpc}.
For example, in~\cite{economic-mpc} the authors present a model predictive control strategy with the objective of minimizing the electricity cost for operating a household heatpump, while maintaining required thermal comfort. This work is extended in~\cite{stochastic-mpc} by setting up a stochastic finite horizon optimization problem for decision making under uncertainty. In another variation of MPC, the authors of~\cite{scenario-mpc} combine a stochastic formulation with a non-linear model of the system. Additionally, works such as~\cite{swiss-mpc, real-world-mpc} present real-world applications of MPC-based controllers. These works show the benefits of using MPC over existing rule-based control and highlight the application procedure used for such real-world implementations. Notably, such implementations have been restricted to office buildings or large commercial spaces. As discussed in~\cite{mpc-challenges, mpc-real-problems}, this is due to the high modeling cost and time required to develop bespoke building models for the MPC. Additionally, MPC-based controllers are highly susceptible to model-bias~\cite{mpc-basic, mpc-bias} and hence a commercial MPC-based solution also requires significant budget for model maintenance. Due to their need for customized models, MPC solutions are cumbersome to deploy at large scale, especially for residential applications. Circumventing these challenges associated with model-based approaches, data-driven model-free control paradigms form an attractive alternative.

Hence, spurred by recent developments in the field of machine learning, data-driven control algorithms have received significant attention, mostly in the form of Reinforcement Learning~(RL)-based controllers~\cite{rl-review}. Works such as~\cite{rl-smart-hem, deep-rl-1} present applications of state-of-the-art RL algorithms for data-driven building controllers. Such RL-based controllers operate in a model-free setting by interacting with the system to create experiences and learning high quality control strategies using these experiences. These works show the potential of using RL-based controllers in the context of building control, demonstrating improvements in performance over existing rule-based approaches. However, while these methods show promising results, challenges such as poor data efficiency, lack of interpretable control logic remain as the major challenges in commercial implementation of such methods~\cite{bems-survey, rl-challenges}.

We identify these shortcomings of MPC-as well as RL-based approaches as a significant gap in existing literature and address them by proposing a physics informed reinforcement learning framework~(PhysQ). Motivated by recent developments in the field of physics informed neural networks~\cite{pinns, pinns-ps}, this PhysQ-based controller uses physics informed neural networks based on prior knowledge related to thermal dynamics of buildings to learn high quality heating control strategies. PhysQ builds on our previous work~\cite{physnet}, where we present a physics informed neural network architecture for control-oriented modeling of buildings. Thus, by effectively using prior building models, we present a RL controller that is data-efficient and builds a control policy using physically relevant features that can be interpreted. Here, we quantitatively assess the performance and data efficiency of PhysQ-based controllers and the role of prior physics knowledge towards the quality of the learnt control policy. The interpretability of the learnt policy is reserved as future work. 

The main contributions of our work are:
\begin{enumerate}
    \item We propose PhysQ, a novel control framework that uses physics-informed representations of the building hidden state and a fitted $Q$-iteration algorithm to obtain high quality control policies~(\secref{sec:physQ}).
    \item In our experiments (\secref{sec:expt-setup}), we show that these PhysQ agents outperform benchmark controllers such as a rule-based controller and a rudimentary MPC~(\secref{subsec:expt1}).
    \item Further, we show that the proposed PhysQ agents are more sample efficient, requiring less training data than conventional FQI agents~(\secref{subsec:expt2})
\end{enumerate}
Before presenting our PhysQ approach in more detail, we position our work against state-of-the-art and state the specific hypotheses underlying our proposition (\secref{sec:related}), as well as the reinforcement learning framework PhysQ is based on (\secref{sec:preliminaries}).

\section{Related Work and Problem Description}
\label{sec:related}
This section presents a non-exhaustive overview of literature related to our work, followed by a summary of the building control problem that we focus on. 

Significant work has been carried out in the context of building control using reinforcement learning~\cite{rl-review}. The most common approach involves modeling the sequential decision making problem as a markov decision process and solving it using a single agent RL algorithm. For eg., works such as\cite{deep-rl-1, rl-smart-hem} present RL based controllers with the objective of optimizing energy consumption in office and residential buildings respectively. Both these studies show that respective RL agents achieve promising results, outperforming baseline controllers by 5-12\%. Differing from such single agent approaches, works such as~\cite{deepmind-hvac-hierarchicalRL, multi-agent-hvac}, propose hierarchical and multi-agent RL strategies for minimizing energy costs in buildings. While these works show that RL-based controllers lead to promising results, there exist significant challenges that need to be addressed~\cite{real-world-rl-discussion}. These include the huge amount of training data required for training, the need for a high-fidelity simulation environment to obtain this data as well as the lack of interpretability of the obtained control logic.

To mitigate these issues, works such as~\cite{mpc-bias, reinforced-mpc} combine RL with model-based control strategies. For eg.,~\cite{mpc-bias} unifies MPC with model-free RL by using the MPC to obtain value estimates. Contrary to this,~\cite{reinforced-mpc} presents a non-linear MPC approach, where model-free value estimates are used in formulation of the receding horizon problem of the MPC. Both these methods show promising results by effectively combining RL and MPC. 

In this work, we present a significantly different approach for combining prior knowledge in model-free RL. We utilize physics informed neural networks~\cite{pinns, physnet} to leverage prior knowledge about system dynamics and learn physically relevant state representations. These physically relevant state representation are then used to better estimate the value functions and learn good control policy~\cite{representation-rl}.

\subsection{Problem Description}
To test the proposed PhysQ framework, we consider a building control problem with the objective of minimizing the cost the energy consumed whilst maintaining user thermal comfort. We focus on a single zone building with a heating source that can be switched ON or OFF by the controller. We assume that the building is exposed to an hourly dynamic price profile that varies every day~\cite{agile-octopus}. To learn a high quality control strategy, the RL controller must learn the thermal behavior of this building given daily changes in external weather conditions. Additionally, the controller must learn to identify suitable heating opportunities by leveraging the volatility of the price. To aid this learning process, the RL-based controller is trained at the end of every day and leverages the forecasts of the price profile, outside air temperature for the next day to learn the control strategy for the next day~\cite{ruelens-FQI}. 

\subsection{Hypothesis}
The proposed PhysQ framework uses prior physics knowledge to obtain physically relevant representations of the hidden states. The physically relevant hidden states along with other observable state components ensure that the $Q$-function obtained is close to optimum, leading to high quality control policies. Thus, with this framework we hypothesize that:
\begin{enumerate}[label=(H{{\arabic*}})]
    \item PhysQ framework with its low dimensional $Q$-function approximations leads to (near) optimal control policies \label{hyp1}
    \item Training such PhysQ agents is data efficient, requiring less training data as compared to conventional FQI approaches \label{hyp2}
\end{enumerate}

\section{Preliminaries}
\label{sec:preliminaries}

\subsection{Markov Decision Process}
We model our sequential decision-making problem as a Markov decision process~\cite{sutton-barto}.
The MDP is defined by its state space~($\mathbf{X}$), action space~($\mathbf{U}$), a state transition function~($f$) and a cost function~($\rho$).\footnote{Conventionally, RL-based agents receive rewards and the agents are trained to maximize the cumulative reward received by the agent. However, in our work we will assume that the objective of the agent is to minimize the cumulative cost. Following this, the discount factor $\gamma$ is set to 1.}
For each discrete timestep $i$, the state evolves as:
\begin{equation}
    \textbf{x}_{i+1} = f(\textbf{x}_{i}, \textbf{u}_{i}, \textbf{w}_{i})
\end{equation}
Here, $w_i \in \mathbf{W}$ represents a random disturbance and models the intrinsic stochasticity of the system.
Additionally, each state transition is associated with a cost term given by 
$c_i = \rho(\textbf{x}_{i}, \textbf{u}_{i}, \textbf{w}_{i})$.
The goal of an agent is to find a policy $\pi : \mathbf{X}\rightarrow \mathbf{U}$ that minimizes the expected $T$-step cost~($J^{\pi}$) starting from a state $\textbf{x}_{0} \in \mathbf{X}$ (\equref{eq:t-step-cost}).
\begin{equation}
    J^{\pi} = \sum_{i=0}^{T}\rho(\textbf{x}_{i}, \pi(\textbf{x}_{i}), \textbf{w}_{i})
\label{eq:t-step-cost}
\end{equation}
This expected cost~($J^{\pi}$) can be expressed as recursive function using a state-action value function, called $Q$-function, defined in \equref{eq:basic-q-function}.
\begin{equation}
    Q^{\pi}(\textbf{x}_{i}, \textbf{u}_{i}) = \E_{\textbf{w}_{i} \in \mathbf{W}} [\rho(\textbf{x}_{i}, \textbf{u}_{i}, \textbf{w}_{i}) + Q^{\pi}(\textbf{x}_{i+1}, \pi(\textbf{x}_{i+1}))]
\label{eq:basic-q-function}
\end{equation}
The optimal $Q$-function, $Q^{*}$, is defined as the best $Q$-function that can be obtained by any policy $\pi$~(\equref{eq:optimum-q}) and satisfies the Bellman optimality equation~(\equref{eq:bellman_eq})~\cite{sutton-barto}.
\begin{align}
    Q^{*}(\textbf{x}_{i}, \textbf{u}_{i}) 
    &\triangleq \min_{\pi} Q^{\pi}(\textbf{x}_{i}, \textbf{u}_{i})
\label{eq:optimum-q}
\\
    &= \E_{\textbf{w}_{i} \in \mathbf{W}} [\rho(\textbf{x}_{i}, \textbf{u}_{i}, \textbf{w}_{i}) + \min_{\textbf{u} \in \mathbf{U}} Q^{*}(\textbf{x}_{i+1}, \textbf{u})]
\label{eq:bellman_eq}
\end{align}
Using the Bellman equation, the optimal policy is computed as:
\begin{equation}
    \pi^{*}(\textbf{x}_{i}) \in \argmin_{\textbf{u} \in \mathbf{U}}Q^{*}(\textbf{x}_i, \textbf{u}).
\label{eq:optimum-policy}
\end{equation}

\vspace{\baselineskip}
We now present a formal description of the various components of the MDP tailored to our building control problem.

\subsubsection{State Description}
Like most real-world systems, the building control problem that we focus on is described by a state that includes some observable components~(e.g., indoor room temperature, outside air temperature) and some components that cannot be observed or measured~(e.g., building thermal mass). The presence of these hidden components (represented as $\textbf{z}$) leads to a partially observable MDP~(POMDP). Due to the hidden states, agents in a POMDP only have access to incomplete state information, making the policy learning process more difficult than the fully observable case. To mitigate this, we include a sequence of past observations to create a state representation, $\textbf{x}^{f}_{i} = \{\textbf{x}^{\text{obs}}_{i-k}, \ldots, \textbf{x}^{\text{obs}}_{i-1}, \textbf{x}^{\text{obs}}_{i}$)\} as presented in~\cite{sutton-barto}. The length of this sequence is referred to as `depth'~($k$) and is a hyperparameter.   

\subsubsection{Actions and Backup Controller}
For our problem, we restrict the action space to a binary setting corresponding to switching the heater ON or OFF. Additionally, we assume an overrule mechanism or a backup controller that guarantees thermal comfort and safety constraints.
The backup control function $\mathcal{B}$, maps the current state~($\textbf{x}_{i}$) and requested action~(${u}_{i}$) to a physical control action $u^{\text{phys}}$~(\equref{eq:backup-general}).
\begin{equation}
     {u}^{\text{phys}}_{i} = \mathcal{B}(\textbf{x}_{i}, {u}_{i}, \textbf{w}_{i}) \quad \forall \ i
\label{eq:backup-general}    
\end{equation}
The settings of this backup controller are unknown to the learning agent and it must learn an optimum policy that accommodates such a backup function.

\subsubsection{Cost Function}
Unlike popular RL approaches that maximize a reward signal, we consider a cost minimization objective: we minimize the cost of energy consumed in one timeslot $i$ of duration%
($\Delta t$,)%
following a dynamic price $\lambda_{i}$.
This cost for timeslot $i$ is defined as:
\begin{equation}
    c_{i} \triangleq \lambda_{i} \> {u}^{\text{phys}}_{i} \> \Delta t
\label{eq:cost-function}
\end{equation}

Since the action space of the learning agent is assumed to be discrete, we select value iteration algorithms due to their data-efficiency as highlighted in~\cite{approx-rl}. Specifically, we focus on fitted Q-iteration~\cite{fqi-ernst}, which is described next.

\subsection{Fitted $Q$-iteration}
Fitted $Q$-iteration~(FQI) is a batch RL technique that iteratively approximates the $Q$-values for a given batch of state transition tuples using the Bellman equation~(\equref{eq:bellman_eq}). As described in~\cite{mpc-vs-rl}, this leads to a series of $Q$-function approximations, $\hat{Q}_{1}, \ldots, \hat{Q}_{N}$, with $N$ being the total number of iterations. Note that, $\hat{Q}_{0}$ is set to $0$ for all state-action pairs. Following~\cite{Ruelens-RL}, we implement the extended FQI algorithm using forecasts of the price profile and other exogenous components such as outside air temperature.
A batch $\mathcal{F}$ comprises four-tuples of the form~$(\textbf{x}_{i}, \textbf{u}_{i}, \textbf{x}_{i+1}, \textbf{u}^{\text{phys}}_{i})$ obtained from previous system interactions. From this batch, we iteratively build a training set $\mathcal{T}$, such that $(\zeta^{l}, o^{l}) \in \mathcal{T}$, for $l=1, 2, \ldots, \lvert\mathcal{F}\rvert$. The target value~$o^{l}$ corresponds to the Q-value of the state-action pair~$\zeta^{l}$, computed using the $Q$-function approximation from the previous iteration. Hence, for the $k^\text{th}$ iteration, $(\zeta^{l}, o^{l})$ are obtained following~\equref{eq:fqi-q-targets}.
\begin{equation}
\begin{split}
    \zeta^{l} &= (\textbf{x}^{l}_{i}, \textbf{u}^{l}_{i})\\
    o^{l} &= \hat{\lambda}_{i} \> \textbf{u}^{\text{phys}, l}_{i} \> \Delta t + \min_{\textbf{u} \in \mathbf{U}}\hat{Q}_{k-1}(\hat{\textbf{x}}^{l}_{i+1}, \textbf{u})
\end{split}
\label{eq:fqi-q-targets}
\end{equation}
A $Q$-function approximator $\hat{Q}_{k}$ is then trained on $\mathcal{T}$.
In \equref{eq:fqi-q-targets}, $\hat{Q}_{k-1}$ represents the function approximator obtained after $(k-1)^{\text{th}}$ iteration, $\hat{\lambda}$ represents the forecasted dynamic price and $\hat{\textbf{x}}^{l}_{i+1}$ represents the modified state representation taking into account forecasts of exogenous variables.
A detailed description of this method can be found in~\cite{Ruelens-RL}.
The total number of iterations~$N$ influences the control horizon of the FQI controller. The building control problem that we consider is an infinite horizon problem. However, a good control policy can be obtained using a horizon of $T$-steps as explained in~\cite{mpc-vs-rl}. To incorporate this $T$-steps horizon, we set the total number of iterations to $2\>T$ and reverse the order of training. This leads to a setting where $Q$-function approximator $\hat{Q}_{2T}$ estimates the expected single step cost and subsequent $Q$-function approximators $\hat{Q}_{k}$ approximate the expected cost over $2\>T-k$ steps. Thus, the series $\hat{Q}_{1}, \ldots, \hat{Q}_{T}$ have a shrinking control horizon from $2\>T$ steps for $\hat{Q}_{1}$ to $T$ steps for $\hat{Q}_{T}$~\cite{shrinking-horizon}. Thus, to obtain the control policy, we only use $\hat{Q}_{1}, \ldots, \hat{Q}_{T}$. 

\begin{algorithm}[t]
\caption{Extended FQI with $2\>T$ distinct function approximators}
\label{alg:back_FQI}
\begin{algorithmic}
\State \textbf{Input}: Batch $\mathcal{F}$, $Q$-function approximators $\hat{\textbf{Q}}=\{\hat{Q}_{1}, \hat{Q}_{2}, \ldots, \hat{Q}_{2T}\}$.
\State \textbf{Initialization}: Set $\hat{Q}_{2T+1}(\textbf{x}, \textbf{u}) = 0 \quad \forall (\textbf{x}, \textbf{u}) \in X \times U$
\For{$k \in [2\>T, 2\>T-1, \ldots, 1]$}
    \State Incorporate forecasts of exogenous variables, prices:
        $$\textbf{x}^{l}_{i+1} \leftarrow \hat{\textbf{x}}^{l}_{i+1}$$
        $$\hat{c}^{l}_{i} = \hat{\lambda}_{k} u^{\text{phys}, l}_{i}\Delta t $$
    \State Create training set $\mathcal{T}$ based on \equref{eq:fqi-q-targets}
    \State \textbf{Regression}: Fit $\hat{Q}_{k}$ on training set $\mathcal{T}$
\EndFor
\State \textbf{Output}: Optimum $Q$-function estimates $\hat{\textbf{Q}}$
\State
\State \textbf{Policy}: Use $\hat{\textbf{Q}} = [\hat{Q}_1, \hat{Q}_2, \ldots, \hat{Q}_T]$ to obtain a greedy policy such that,
\begin{equation}
    u^{*}_{t} = \argmax_{u \in U} \hat{Q}_{t}(\textbf{x}_{t}, \textbf{u}) \qquad \forall t \in [1, 2, \ldots, T]
\label{eq:back_FQI_policy}
\end{equation}
\end{algorithmic}
\end{algorithm}

\section{PhysQ Framework}
\label{sec:physQ}
Building on our previous work~\cite{physnet}, we propose a two-step algorithm that leverages prior physics knowledge to obtain optimal control policies. Previously, works such as~\cite{pomdp-belief, lr-one-objective} have explored the role of learnt hidden representations in learning good control policies and the use of different objectives to learn such representations. However, the representations are learnt implicitly and are not constrained to follow a particular behavior. Differing from such approaches, we explicitly learn representations that closely resemble the physical behavior of the hidden states of the system. The following subsections detail the architecture used and the training algorithm followed in our PhysQ framework.

\begin{figure}
    \centering
    \includegraphics[width=0.45\linewidth]{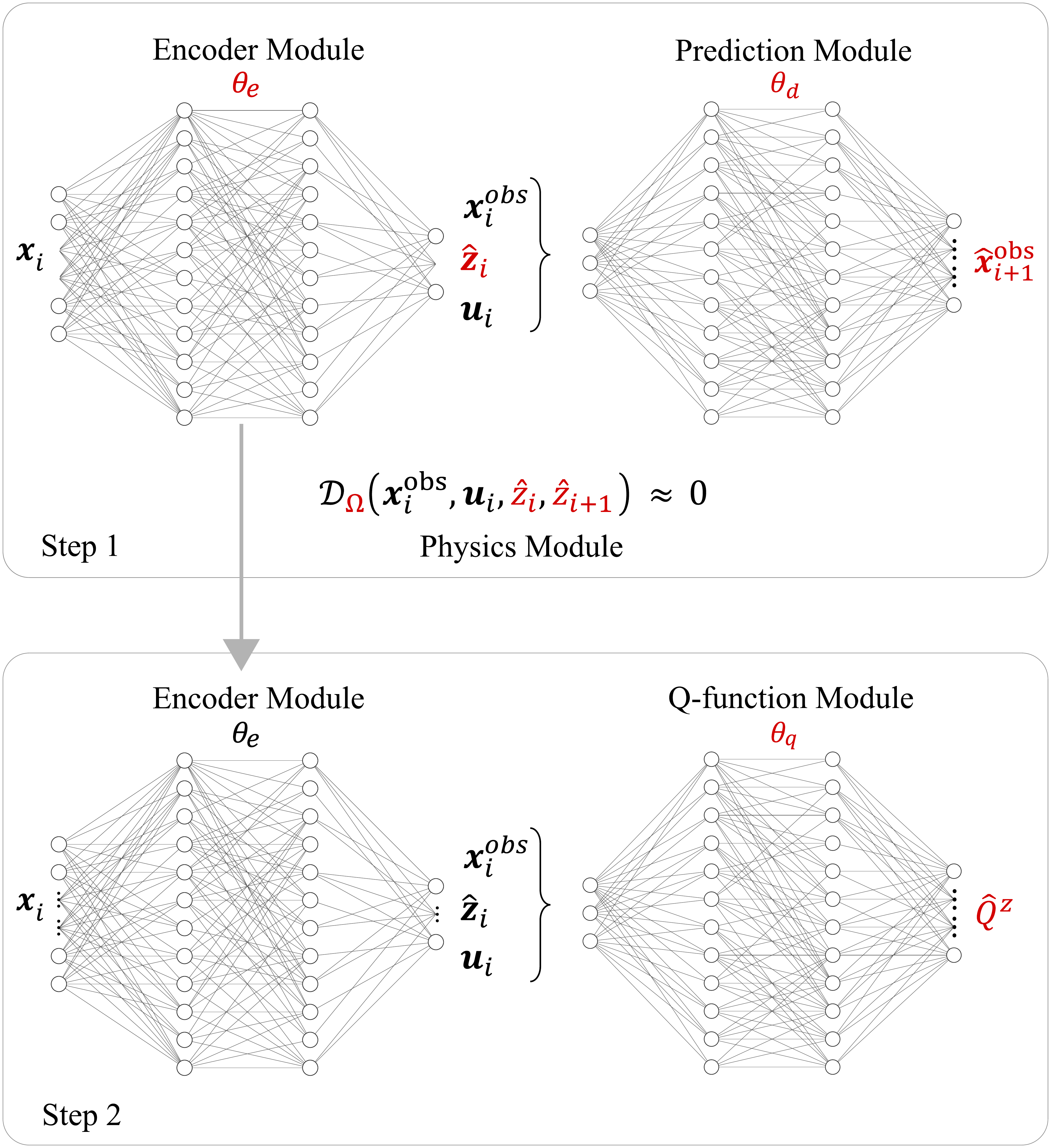}
    \caption{Proposed PhysQ Architecture: In Step one~(top), a physically relevant representation of the hidden state is learnt. Step 2~(bottom) represents learning the $Q$-function. The encoder module from Step 1 is used directly in Step~2, as indicated by the arrow. The predicted quantities and learnt parameters from each step are highlighted in red. This block represents a single function approximator and is repeated $2T$ times as discussed in \algoref{alg:back_FQI}.}
    \label{fig:physq}
\end{figure}

The key idea of the PhysQ architecture is to use an encoder module to obtain a physically relevant, low-dimensional representation of the hidden state, which is designed to support next-state prediction, while approximately adhering to a set of differential equations of the physical model of the system.
In Step~1, we train this Encoder module together with an auxiliary Prediction module, where the Physics module represents the physics model of the system. Then, in Step~2, we train the $Q$-function defining our policy, based on the physically relevant representations of the system state along with other observed state components. Below, we detail the neural network models for both steps as sketched in \figref{fig:physq}, as well as the training procedure outlined in \algoref{alg:physQ_train}. Note that, we use $2\>T$ different function approximators, each using the architecture illustrated in \figref{fig:physq}.

\subsection{Architecture}

\subsubsection{Step 1} \textit{Extracting physically relevant hidden states.}
The first step uses the physics informed neural network from~\cite{physnet}, comprising an encoder module (with learnable parameters~$\theta_e)$, a prediction module (with learnable parameters~$\theta_d$) and a physics module~(with learnable parameters~$\Omega$).
This encoder-based network uses input state~($\textbf{x}_{i}$) to obtain a physically relevant representation of the hidden state~($\hat{\textbf{z}}_{i}$).
The objective for the latter hidden state is to adhere to a physics-based model as good as possible (\cf the $D_\Omega(.) \approx 0$ constraint).

\subsubsection{Step 2}\textit{Low-dimensional $Q$-function and optimal policy.}
In this step, the trained encoder module from Step~1 is combined with a $Q$-function module~($\theta_q$) that estimates a low-dimensional $Q$-function~($\hat{Q}^{z}$). The input space for this low-dimensional $Q$-function includes the observable state components ($\textbf{x}^{\text{obs}}_{i}$), encoder output~($\hat{\textbf{z}}_{i}$) and action~($\textbf{u}_{i}$). An optimal action~($\textbf{u}^{*}_{i}$) is then computed using this as:
\begin{equation}
    \textbf{u}^{*}_{i} = \argmin_{\textbf{u} \in U} \hat{Q}^{z}(\textbf{x}^{\text{obs}}_{i}, \hat{\textbf{z}}_{i}, \textbf{u})
\label{eq:step2-policy}
\end{equation}

\subsection{Training}
\begin{algorithm}[t]
\caption{PhysQ training procedure}
\label{alg:physQ_train}
\begin{algorithmic}
\State \textbf{Input}: Set of state-action tuples $\mathcal{F}$, PhysQ agent parameterized by $\{ \theta_e, \theta_d, \Omega, \hat{\textbf{Q}^{z}}=\{\hat{Q}^{z}_{1}, \hat{Q}^{z}_{2}, \ldots, \hat{Q}^{z}_{2T}\} \}$.
\State \textbf{Initialization}: Set $\hat{Q}_{2T+1}(\textbf{x}, \textbf{u}) = 0 \quad \forall (\textbf{x}, \textbf{u}) \in X \times U$
\State \textbf{Step 1:} Train encoder module~($\theta_{e}$):
        \State Training set $\mathcal{T}=\{ (\textbf{x}^{l}_{i}, \textbf{u}^{l}_{i}), (\textbf{x}^{\text{obs}, l}_{i+1}) \}_{l=1}^{\#\mathcal{F}}$
        \State Perform regression on $\mathcal{T}$ using \equref{eq:physnet-eq}
\State Freeze encoder module $\theta_e$
\State Create modified batch~($\mathcal{F'}$) 
$$\mathcal{F'}=\{( (\textbf{x}^{\text{obs},l}_{i}, \hat{\textbf{z}}^{l}_{i}), \textbf{u}^{l}_{i}, (\textbf{x}^{l}_{i+1}, \hat{\textbf{z}}_{i+1}), \textbf{u}^{\text{phys}, l}_{i})\}_{l=1}^{\#\mathcal{F}}$$
\State \textbf{Step 2:} Train $Q$-function Modules:
\quad    \State Use modified batch $\mathcal{F'}$
\quad    \For{$k \in [2T, 2T-1, \ldots, 1]$}
        \State Incorporate forecasts of exogenous variables, prices:
            $$\textbf{x}^{l}_{i+1} \leftarrow \hat{\textbf{x}}^{l}_{i+1}$$
            $$\hat{c}^{l}_{i} = \hat{\lambda}_{k} \textbf{u}^{\text{phys}, l}_{i}\Delta t $$
        \State Create training set $\mathcal{T}$ based on \equref{eq:fqi-q-targets}
        \State \textbf{Regression}: Fit $\hat{Q}^{z}_{k}$ on training set $\mathcal{T}$
    \EndFor
\State \textbf{Output}: Optimum $Q$-function estimates $\hat{\textbf{Q}}^{z}$
\State
\State \textbf{Policy}: Use $\hat{\textbf{Q}}^{z} = [\hat{Q}^{z}_1, \hat{Q}^{z}_2, \ldots, \hat{Q}^{z}_T]$ to obtain a greedy policy such that,
$$    \textbf{u}^{*}_{t} = \argmin_{\textbf{u} \in U} \hat{Q}^{z}_{t}(\textbf{x}^{\text{obs},l}_{t}, \hat{\textbf{z}}^{l}_{t}, \textbf{u}) \qquad \forall t \in [1, 2, \ldots, T]$$
\end{algorithmic}
\end{algorithm}
In Step~1, the encoder module is trained in a supervised manner as a prediction model to learn the dynamics of the building and extract the hidden states.
Consequently, the encoder module is trained once at the beginning and then shared across all approximators.
For training the encoder module, the loss functions used are given by \equref{eq:physnet-eq}.
\begin{equation}
\begin{split}
        \text{Loss} &= \mathcal{L}_{\text{pred}} + \mu \> \mathcal{L}_{\text{phys}} \\
    \mathcal{L}_{\text{pred}} &= \frac{1}{N} \sum_{i=1}^{N} \left(\, \textbf{x}_{i+1}^{\text{obs}} - \hat{\textbf{x}}_{i+1}^{\text{obs}}\,  \right)^{2}\\
    \mathcal{L}_{\text{phys}} &= \frac{1}{N} \sum_{i=1}^{N} \left( \, \mathcal{D}_{\Omega}^{\phantom{b}} (\, \textbf{x}_i, \textbf{u}_i, \hat{\textbf{z}}_{i}, \hat{\textbf{z}}_{i+1}, \textbf{w}_{i}) \, \right)^{2} \\
\label{eq:physnet-eq}
\end{split}
\end{equation}
Following the training of the encoder in Step~1, hidden states for all state-action pairs in batch~$\mathcal{F}$ are calculated to create a new batch that is then used in Step~2 to train the $Q$-function module in an iterative manner.
The output is a low-dimensional $Q$-function approximation which defines our learnt policy.

\section{Experiment Setup}
\label{sec:expt-setup}
To test the performance of our proposed PhysQ framework and validate our hypotheses, we performed a set of experiments using a single zone building simulator model. We detail the simulation environment used, the type of experiments performed and other experiment parameters. 

\subsection{Building Simulator}
A simplified scenario is considered with a single-zone building heated using a single heat source.
The building model is based on the two-state RC model presented in~\cite{Vrettos2018}.
The environment states include the room temperature~($T_{r}$) that is observed by the agent and the temperature of the building thermal mass~($T_{m}$) which is hidden. Additionally, we only consider the impact of outside air temperature~($T_a$) on the environment states.

\subsubsection{Backup Controller}
The backup controller described in \secref{sec:preliminaries} is designed with a user comfort band of $\pm 2$\textdegree{C} around a desired temperature of $20$\textdegree{C} and modelled as:
\begin{equation}
    u^{\text{phys}}_{i} =   \begin{cases}
                                0       &T_{r,i}        > 22\text{\textdegree{C}} \\
                                u_{i}\>u^{\text{max}}   &18 \text{\textdegree{C}} \leq T_{r,i} \leq 22\text{\textdegree{C}}\\
                                u^{\text{max}} & T_{r,i} < 18\text{\textdegree{C}}\\
                            \end{cases}
\label{eq:backup}.
\end{equation}

We further assume the backup controller will check the room temperature $T_{r,i}$ every minute, and thus override the current action $u_i$ as soon as $T_{r,i}$ goes out of the assumed [18\textdegree{C}, 22\textdegree{C}] comfort bounds. Further, in case of such overriding backup action $u^{\text{phys}}_{i}$, we will maintain it for the remainder of timeslot~$i$, to avoid switching the heat source too frequently.

\subsubsection{Dynamic Energy Prices}
\begin{figure}
    \centering
    \includegraphics[width=0.65\linewidth]{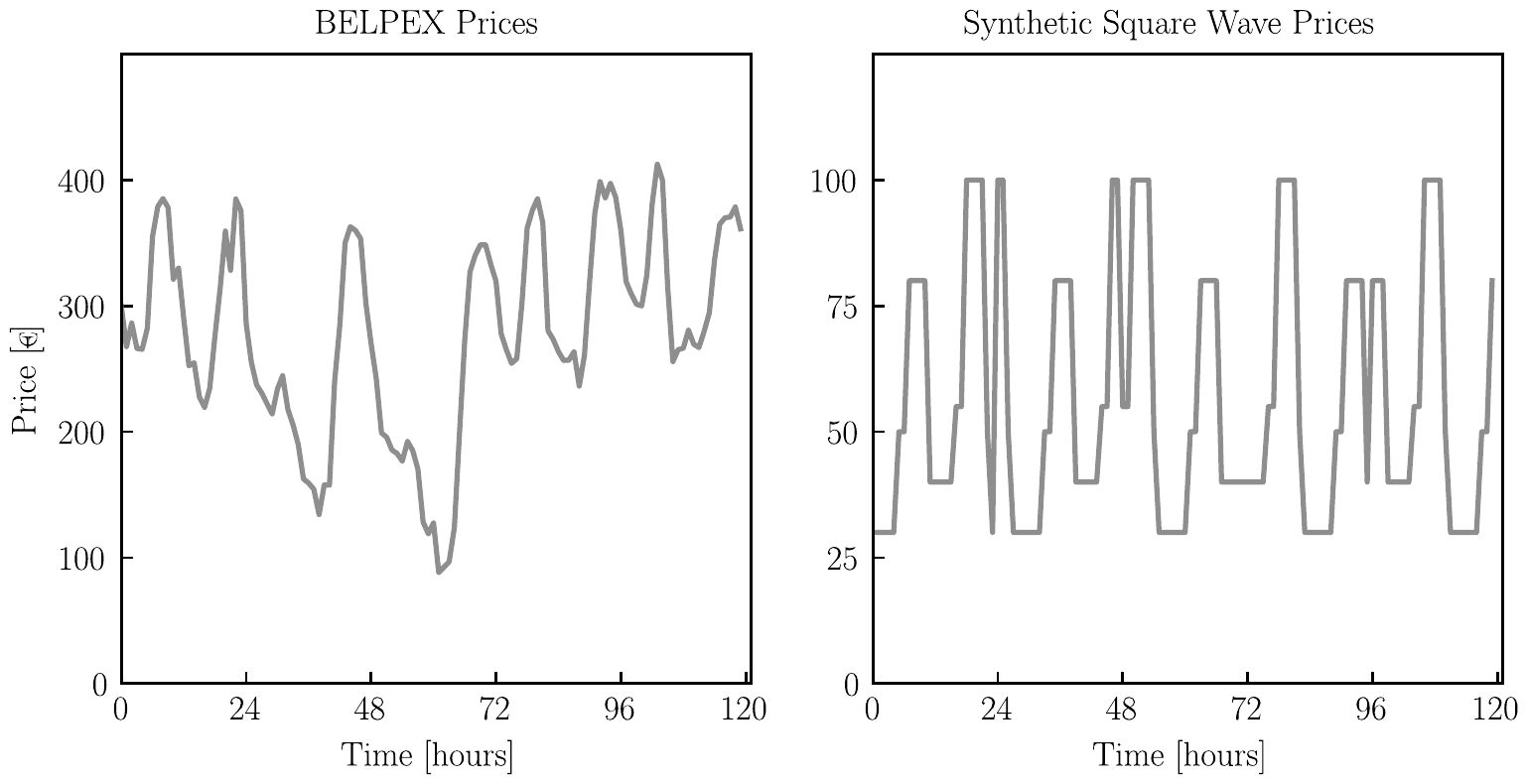} 
    \caption{Different price profiles used during simulations. The graph on left illustrates the synthetic square wave price and the one on right shows the BELPEX day-ahead price for a sample set of 5~days. These price profiles are a part of the test environment that was held-out while training and used for validating the performance of different agents.}
    \label{fig:price-profiles}
\end{figure}

As presented in \secref{sec:preliminaries}, the agent learns to minimize the cost of energy following a dynamic price profile. We assume that the forecast of this price are available in advance\footnote{In particular, we assume the agent will be trained once nightly, based on the prices received for the next day.} and is provided to the agent following~\algoref{alg:back_FQI}.
To analyze how well a trained agent responds to the time-varying price signal, we experiment with two types of price profiles: 
\begin{enumerate*}[label=(\roman*)]
    \item Synthetic square wave prices; 
    and
    \item BELPEX day-ahead prices~\cite{entsoe-belpex};
\end{enumerate*}
as shown in \figref{fig:price-profiles}.
The synthetic square wave prices were designed with clear price peaks and troughs that were shifted in time.
This aided the learning process of the agents and enabled us to analyze the robustness of polices over time-shifted price extrema.
On the other hand, the BELPEX price profile are real prices and allow us to assess how well the agents can perform in a real-world price scenario.

\subsection{RL Agent Configurations}
As described in \secref{sec:preliminaries}, the agents observe a few state components such as room temperature, outside air temperature and need to take a cost minimizing action based on these observations, leading to a POMDP. Hence, the agent inputs include a series of past room temperature observation of length $k$. The length of this sequence controls the amount of past information given to the agent and compensates for the hidden states. For our problem, we set $k=$ 4, providing the agent wit room temperature information over the past 4~hours.

We compare our PhysQ agents with other conventional FQI agents that use different types of function approximators: Extremely randomized (EXTRA) trees~\cite{extra-trees} and neural networks. These represent agents commonly used in literature and have shown promising results with respect to their performance and training stability~\cite{Ruelens-RL, mpc-vs-rl}. The hyperparameters for these function approximators are presented in \appref{app:hyperparameters}.
Considering the dynamics of the building, we set the control horizon to 24 hours, leading to 48 distinct instances of the $Q$-function approximator as presented in \algoref{alg:back_FQI}.
Thus, to learn the policy for a given day, the agent uses a forecast for that day and the next, of day-ahead prices and outside air temperature. We assume that this forecast is available and accurate, with incorporation of forecasting inaccuracies reserved as a future topic.  

\subsection{Prior Physics Knowledge}
As detailed in \secref{sec:physQ}, the PhysQ framework uses prior physics knowledge related to the system to learn a meaningful representation of the hidden states of the system.
For our experiments, we assume that the building has one hidden state, namely the thermal mass temperature.
To extract this, we follow the approach presented in~\cite{physnet} and provide the PhysQ agents with the following first-order model:
\begin{equation}
\begin{split}
    T^{\mathcal{M}}_{r,i+1} &= a_{11} T_{r,i} + a_{12} \hat{T}_{m,i} + b_{1}u^{\text{phys}}_{t} + c_{11} T_{a,t} \\
    T^{\mathcal{M}}_{m,i+1} &= a_{21} T_{r,i} + a_{22} \hat{T}_{m,i}\\
\end{split}
\label{eq:building-physics}.
\end{equation}
Here, $a_{ij}$, $b_{1}$, $c_{11}$ represent the building parameters (\ie $\Omega$ from \figref{fig:physq}) and $\hat{T}_{m,i}$ represents the encoder module output (noted as $\hat{\textbf{z}}_{i}$ in \figref{fig:physq}).
Thus, the encoder module of PhysQ is trained using \equref{eq:building-physics} and \equref{eq:physnet-eq}.

\subsection{Training Strategy}
For training an RL agent, we followed two distinct strategies:
\begin{enumerate*}[label=(\roman*)]
    \item \label{enum:growing_batch} Growing Batch Strategy, and 
    \item \label{enum:fix_batch} Fixed Batch Strategy.
\end{enumerate*}
In \ref{enum:growing_batch}, an agent starts with an empty state transition batch~$\mathcal{F}$, uses an $\epsilon$-greedy exploration strategy based on its current policy to add sample samples to this batch and retrain its policy periodically using \algoref{alg:back_FQI} on this growing batch to improve the policy. In the $\epsilon$-greedy exploration strategy, the agent selects a random, exploratory action with a probability of $\epsilon$ and a greedy action with a probability of $1-\epsilon$~\cite{sutton-barto}. We initially set $\epsilon=$ 0.6 and after every day reduce it by $9\%$.

To speed up computations, the agents were trained after every 5 days (instead of daily training) using the batch of data collected till that time. While this mimics a real-world deployment scenario, comparing different agent is difficult as each agent trains on a different batch of data due to the random nature of exploratory actions. Hence, to compare different agents and draw any conclusions related to the advantages of using a type over the other, we used~\ref{enum:fix_batch}. Here, batches of different sizes based on the number of training days encountered were created a priori using a default RL agent following~\ref{enum:growing_batch} and different types of agents were trained on these fixed batches. This allowed us to compare the performance of different agents that have been trained on identical data and the results represent a more fair comparison of the quality of each regression technique.

\subsection{Training and Test Environment}
We used a separate training and test environments to assess the performance of each agent.
Both these environments have identical building parameters and differ in the values of exogenous parameters such as day-ahead prices and outside air temperature.
To create the test environment, we sample and hold 
out a subset of price profiles and outside air temperature profiles equivalent to 5 days.
The held-out price profiles corresponding to the 5 test days are shown in \figref{fig:price-profiles}.

The remaining data, spanning 30 days, is given to the training environment.
This ensures that the exogenous parameters take values from the same distribution but are not seen by the agent during training.
The 5 test days are used to compare the performance of different agents using the cumulative 5-day cost as a comparison metric.
We train 5 instances of each agent type and report the mean and standard deviations of this 5-day cost.
Additionally, while following the fixed batch training strategy, each instance of a given agent type is trained on a unique batch of data, thus allowing for variations in generated batch as well as different random initialization of the agents.
The following section presents the results of the experiments that were performed. 

\subsection{Benchmark Controllers}
Besides different types of FQI agents, we compare the performance of our PhysQ approach against a set of benchmark controllers that include: Business-as-usual~(BAU) controller and MPC. The BAU represents a simple rule-based controller resembling the backup controller modelled in \equref{eq:backup}. Such a controller solely focuses on the thermal comfort of the user and is insensitive to the price. The MPC benchmark is a simple MPC based on the building model described in \equref{eq:building-physics}. The MPC benchmark uses the same action space and time horizon as the RL agents along with an accurate model of the building. We consider two different control frequencies for the MPC (hourly and quarterly). More details of the MPC are presented in Appendix \ref{app:mpc}. 

\section{Results}
\label{sec:results}

\subsection{Experiment~1: PhysQ Validation}
\label{subsec:expt1}
\begin{figure}[t]
\centering
    \includegraphics[width=.6\linewidth]
{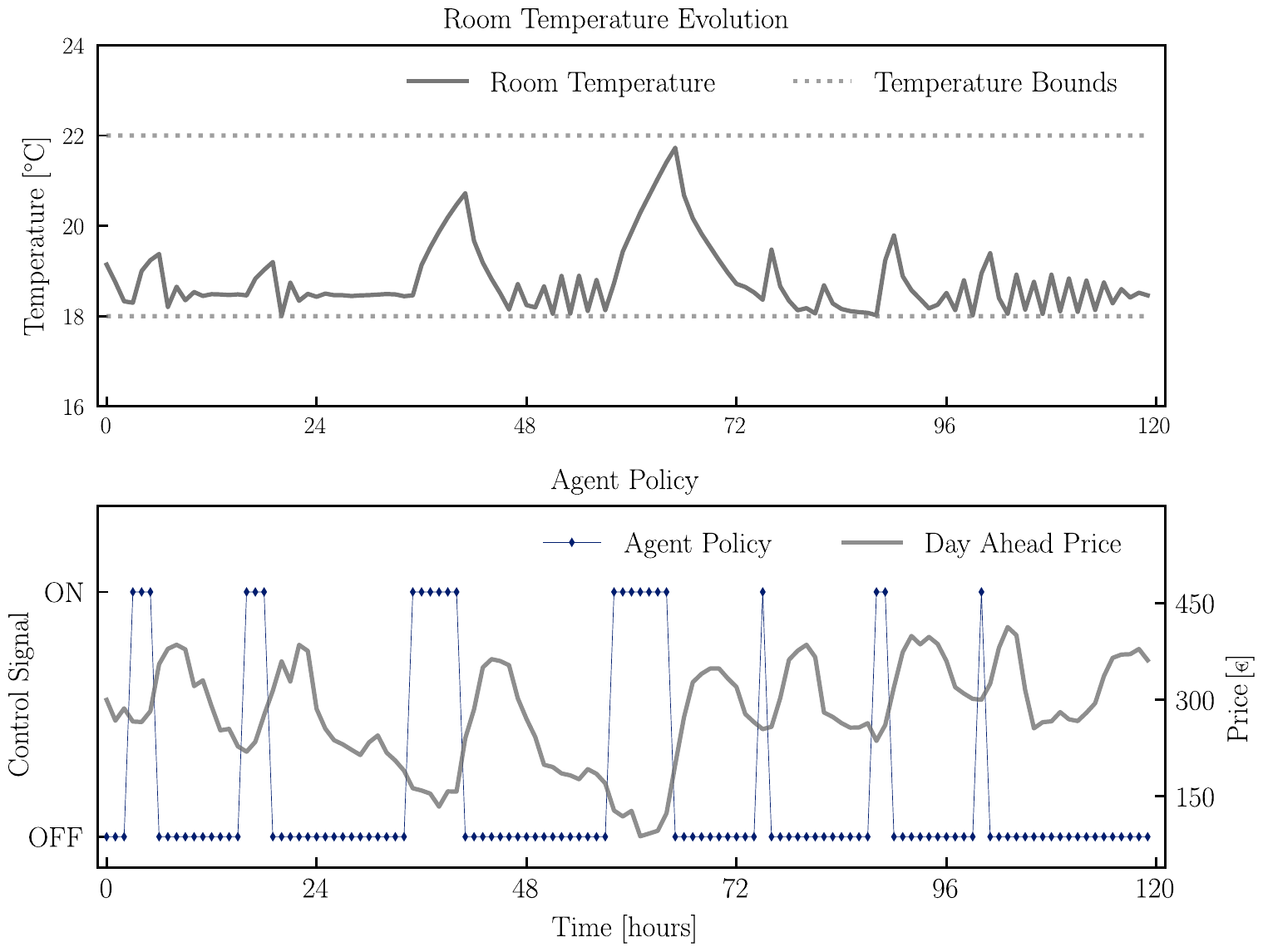}
\caption{Simulation results illustrating the behavior of a trained PhysQ agent in the test environment for the BELPEX price scenario. The top row shows the variations in indoor room temperature of the building caused by following the agent policy. The bottom row shows the actions taken by the agent at different timesteps.}
\label{fig:expt1-belpex-state-action}
\end{figure}

We focus on validating hypothesis~\ref{hyp1}, verifying that the proposed PhysQ architecture can learn good policies for the given building thermal problem. The obtained policy is compared with the benchmark controllers, with the MPC being an hourly MPC. We train the PhysQ agents following the growing batch training strategy for a total of 30 days (24 samples for each day) for both types of price profiles (square wave and BELPEX). \Figref{fig:expt1-belpex-state-action} presents the agent interactions in the test environment, showing the variations in room temperature that result by following the learnt policy in the BELPEX price scenario. 

The learnt policy in \figref{fig:expt1-belpex-state-action} indicates that the PhysQ agent manages to determine suitable~(\ie cheap) time intervals for heating. By switching ON at times when the price is low and remaining OFF when the price is high, the agent manages to minimize the cost of energy consumption while maintaining the required comfort levels. Further, this policy superimposed with the backup controller~(\equref{eq:backup} led to a cost saving of about $~9\%$ over a business-as-usual controller, demonstrating a significant potential for reducing energy bills. \Tabref{tab:expt-1} summarizes the results obtained by the PhysQ agent for different price scenarios along with the benchmark controllers~(BAU), MPC. The total cost over the 5 validation days is used as a comparison metric.  

\begin{table}[t]
    \centering
    \caption{Comparison of PhysQ agent policy with benchmark controllers.}
    \label{tab:expt-1}

    \begin{tabular}{ccc}
        \toprule
        Price Scenario                                  & Controller    & Cost\\ 
        \midrule
        \multirow{3}{*}[-5pt]{Synthetic Square Wave}    & PhysQ         & \texteuro\,4,520 \\
                                                \cmidrule{2-3}
                                                        & BAU           & \texteuro\,5,589 \\
                                                \cmidrule{2-3}
                                                        & Binary MPC    & \texteuro\,4,620\\
        \midrule
        \multirow{3}{*}[-5pt]{BELPEX}                   & PhysQ         & \texteuro\,25,758 \\
                                                \cmidrule{2-3}
                                                        & BAU           & \texteuro\,28,214 \\
                                                \cmidrule{2-3}
                                                        & Binary MPC    & \texteuro\,28,152\\
        \bottomrule
    \end{tabular}
\end{table}

This experiment demonstrated that the proposed PhysQ agent can learn good control policies, thus validating hypothesis~\ref{hyp1}. The trained PhysQ agent outperforms benchmark controllers such as the BAU and the binary MPC. Note that the Binary MPC takes actions at an hourly frequency which leads to a conservative policy, where the apartment is heated longer to ensure that the room temperature never falls below the comfort setting in between two consecutive actions. In contrast the PhysQ policy is a combination of the backup controller with a cost minimization policy. This combination leads to purely cost minimization actions by the agent such that the backup control is activated frequently for shorter time duration to ensure user comfort. This leads to a combined policy with a finer control resolution~($<$~1~hour) as compared to the binary MPC, resulting in a superior performance. Consequently, for subsequent experiments we use an MPC-based controller with a control frequency of 15 minutes to as a benchmark `optimal' policy. 

\subsection{Experiment~2: Performance across different training sizes}
\label{subsec:expt2}
\begin{figure*}[t]
\centering
    \includegraphics[width=\textwidth]{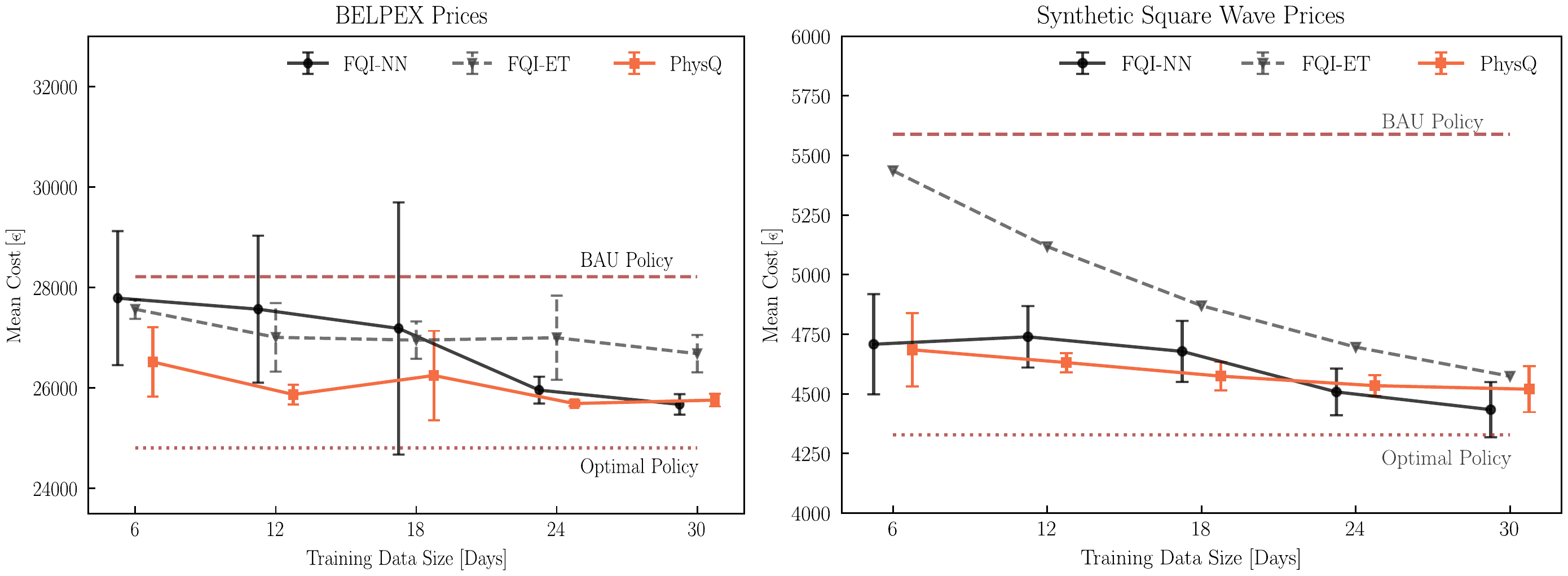}
\caption{Simulation results comparing the performance of different FQI agents for increasing training data sizes. We show the mean performance across the 5 instances along with error bars indicating the standard deviation. Additionally benchmarks including BAU and optimal policy are indicated using dotted lines.}
\label{fig:fig:expt2-performance}
\end{figure*}

We now test hypothesis~\ref{hyp2} and investigate the training data requirements of the PhysQ agent, compared to other FQI approaches. We follow the fixed batch training strategy, using batches of 5 different sizes, varying from 6 to 30 days, each day comprising of 24 hourly samples. For each agent type, 5 instances of the agent were trained. Each instance was trained on a different set containing batches of every considered size. This ensured that the comparison metrics obtained were less sensitive to the quality of the batch and representative of the true performance of the agent type selected.
\figref{fig:fig:expt2-performance} compares our PhysQ agent's performance to that of FQI both using EXTRA Trees~(FQI-ET) and Neural networks~(FQI-NN). The graphs show the mean cost values (summed over the 5 test days), with error bars indicating the standard deviation over the 5~runs per agent type. 

These simulation results show that when the size of training data is small, the PhysQ agents outperform the other two types of agents. With increasing training data size, FQI-NN achieves a similar performance as PhysQ.
This figure also shows that for all training data scenarios, the PhysQ agents outperformed the BAU policy while performing close to the optimum policy. This optimum policy is based on a quarterly MPC benchmark controller described in Appendix~\ref{app:mpc}. Further, the small spread of the error bars indicates that even with less training data, PhysQ agents achieve similar performance when trained using different weight initialization and training batches. This indicates that the PhysQ's training performance is more robust than FQI-NN, which is more susceptible to training instability. From this experiment we conclude that the proposed PhysQ agents require less training data and lead to stable policies that outperform other conventional FQI agents, validating our second hypothesis~\ref{hyp2}. 

\subsection{Experiment 3: Ablation Study for Physics Quality}
The previous experiment showed that the PhysQ agents leverage prior physics knowledge to learn good control policies in a sample efficient manner.
We now perform an ablation study to ascertain the role of this prior physics knowledge in the policy obtained. We thus aim to validate that PhysQ's improved performance does not result from a regularizing effect that may be achieved by any constraining $D_\Omega$ model (see \figref{fig:physq}), but from a physically relevant one. Specifically, we investigated the impact of providing wrong prior knowledge to the PhysQ agent on the learnt policies.

As detailed in \secref{sec:physQ}, the prior physics that we used for PhysQ is given by a system of linear equations,~\equref{eq:physnet-eq}, used to train the encoder module to extract physically relevant latent representations of the building thermal mass. To perform this ablation study, we change the true building physics equations to a wrong set of equations to define $D_\Omega$,
given by:
\begin{equation}
    T^{\mathcal{M}}_{m,t} = \lceil T_{r,t} \rceil
\label{eq:bad_physics},
\end{equation}
where $\lceil \cdot \rceil$ represents the ceiling operation and models the temperature of the building thermal mass as a step function in phase with the indoor room temperature of the building. This significantly differs from the real behavior of the building thermal mass, which is an inertia quantity that will be out-of-phase with the room temperature variations. Thus, the $Q$-function (and hence the control policy) will be learnt using an incorrect representation of the building thermal mass.

\begin{figure*}[t]
\centering
    \includegraphics[width=\textwidth]{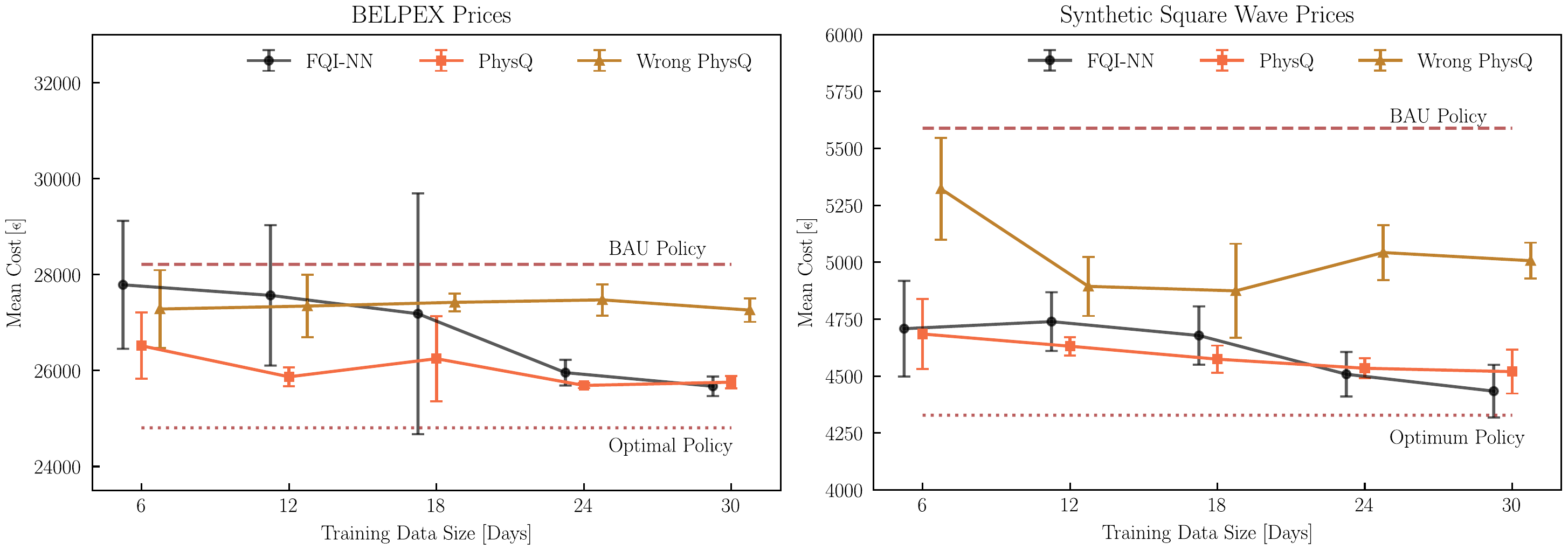}
\caption{Simulation results comparing the quality of physics in the PhysQ framework. Figure on left shows results for the synthetic square wave scenario and on right for the BELPEX prices. The results present the mean cost over the 5 instances along with error bars indicating standard deviation. The performance is benchmarked by BAU and optimal policy shown using dotted lines and FQI-NN agent shown in grey.}
\label{fig:expt3-physics-quality}
\end{figure*}

We follow the same comparison strategy as the previous experiment and compare this `wrong PhysQ' with the correct PhysQ agents, benchmarked using FQI-NN, BAU policy and Optimal policy.
\Figref{fig:expt3-physics-quality} present the results of this experiment for both price scenarios.
These results show a clear performance gap between the `wrong PhysQ' agents that use incorrect physics priors versus the correct PhysQ agents.
This performance gap illustrates the role of prior physics knowledge in the proposed PhysQ framework and suggests that accurate physics information is a key requirement for obtaining good control policies. 

\section{Conclusion and Future Work}
We presented the PhysQ framework, a novel method leveraging physics informed neural networks that extract physically relevant representations of the hidden state of the system and use them to learn a low-dimensional $Q$-function to obtain optimal control policies.
Our experiments demonstrated that PhysQ based agents learnt better control policies compared to a rule-based business-as-usual control strategy, leading to monetary savings of about 9\%.
Further, the PhysQ agents required fewer training samples to learn such policies compared to conventional FQI agents, validating our hypothesis related to the sample efficiency of such agents.
These experiments also highlighted to the role of prior physics knowledge towards the final policy learnt by the agents.
With this, we show that by incorporating prior knowledge, data-driven controllers can utilize data more efficiently, making them a feasible solution for real-world deployments.

We tested the PhysQ framework using an in-house simulator model. Future work will focus on real-world settings. Thus, we will assess the performance of PhysQ for managing the heating patterns of real buildings. This work includes two main topics: (i)~implementation of PhysQ on real buildings; and (ii)~analyzing the impact of forecasting errors in exogenous variables on the performance of PhysQ agents. Besides real-world experimentation, other future direction includes work related to the interpretability of the policy obtained using PhysQ. This includes exploring techniques such as policy distillation to obtain intuitive control policies that can be easily explained to end-users. 

\section*{Funding Information}
This work was supported by the European Union's Horizon 2020 research and innovation programme under the projects BRIGHT (grant agreement no.\ 957816), RENergetic (grant agreement no.\ 957845) and BIGG (grant agreement no.\ 957047).

\appendix
\section{Appendix}
\subsection{Hyperparameters}
\label{app:hyperparameters}
This section lists the hyperparameters chosen for each of the agent type. Neural networks were implemented using PyTorch~\cite{pytorch} and EXTRA trees using scikit-learn~\cite{scikit-learn}. 

\begin{table}[t]
\centering
\caption{Hyperparameters for PhysNet Architecture}
\begin{tabular}{l c}
    \toprule
     Parameter & Value\\
    \midrule
        Optimizer & Adam\\
        Learning Rate & 0.001\\
        Activation Function & $\text{ReLu}$\\
        Batch Size  &   2,048 \\
    \midrule
    \multicolumn{2}{c}{\textit{Encoder Module} ($\theta_{e})$} \\
    \midrule
        Hidden Layers & [32, 32]\\
    \midrule
    \multicolumn{2}{c}{$Q$-\textit{function Module} ($\theta_{q}$)} \\
    \midrule
        Output Size & 2 \\
        Hidden Layers & 2\\
        Neurons per layer & 32\\
    \midrule
    \multicolumn{2}{c}{\textit{Dynamics Module} ($\theta_{d}$)} \\
    \midrule
        Hidden Layers & [128]\\
    \bottomrule
\end{tabular}    
\label{tab:hp-physQ}
\end{table}

\begin{table}[t]
    \centering
    \caption{Hyperparameters for FQI-NN}
    \begin{tabular}{l c}
    \toprule
     Parameter & Value\\
    \midrule
        Optimizer & Adam\\
        Learning Rate & 0.01\\
        Activation Function & $\text{ReLu}$\\
        Batch Size  &  2,048\\
        Output Size & 2 \\
        Hidden Layers & [48, 48]\\
    \bottomrule
    \end{tabular}
\label{tab:hp-fqi-nn}
\end{table}

\begin{table}[t]
    \centering
    \caption{Hyperparameters for FQI-ET}
    \begin{tabular}{l c}
    \toprule
     Parameter & Value\\
    \midrule
       \# of Estimators & 100\\
        Minimum Sample Split & 3\\
        Minimum Samples Leaf &   1\\
    \bottomrule
    \end{tabular}
\label{tab:hp-fqi-et}
\end{table}

\subsection{Model Predictive Control}
\label{app:mpc}
To benchmark different RL agents, we employ a model predictive control (MPC) approach presented in~\equref{eq:mpc-opt}.
Here, $D_{\Omega}$ refers to the model of the system, given by~\equref{eq:building-physics} in our case. $\textbf{x}_{t}$ represents the states of the system, which in our case are $T_{r,t}$ and $T_{m,t}$. We assume binary decision actions $u_t \in \{0, u_{\text{max}}\}$. We consider two different control timesteps: $\text{MPC}_{\text{h}}$ considers hourly control frequency~($T_{\text{MPC}}=24)$) and $\text{MPC}_{\text{q}}$ that takes actions every 15 minutes~($T_{\text{MPC}}=96)$). Note that, \tabref{tab:expt-1} presents the results of $\text{MPC}_{\text{q}}$ and the optimal policy indicated in \figsref{fig:fig:expt2-performance}{fig:expt3-physics-quality} is based on $\text{MPC}_{\text{h}}$. The user comfort constraints are set using~\equref{eq:backup}. 
\begin{equation}
    \begin{split}
        & \min_{u_{1}, \ldots, u_{T_{\text{MPC}}}} \sum_{t=1}^{T_{\text{MPC}}} \lambda_{t} u_{t} \\
        &\text{s.t.} \ \textbf{x}_{t+1} = D_{\Omega} (\textbf{x}_{t}, u_{t})\\
        & \quad \ \  \textbf{x}_{\text{min}} \leq \textbf{x}
        _{t} \leq \textbf{x}_{\text{max}},\\
        & \quad \ \  u_{t} \in  [0, u_{\text{max}}], \ \forall t \\
\end{split}
\label{eq:mpc-opt}
\end{equation}

\bibliography{bibliography_bib.bib}{}
\bibliographystyle{IEEEtran}

\end{document}